\documentclass[%
reprint,
showpacs,preprintnumbers,
 amsmath,amssymb,
 aps,
]{revtex4-1}

\usepackage{graphicx}
\usepackage{dcolumn}
\usepackage{bm}
\usepackage{subfigure} 
\usepackage{color,soul}
\usepackage{ulem}

\begin{document}


\title{Size-dependent contact angle, and wetting and drying transition of a droplet adsorbed on to a spherical substrate: Line tension effect}

\author{Masao Iwamatsu}
\email{iwamatsu@ph.ns.tcu.ac.jp}
\affiliation{%
Department of Physics, Faculty of Liberal Arts and Sciences, Tokyo City University, Setagaya-ku, Tokyo 158-8557, Japan
}%



\date{\today}

\begin{abstract}
The size-dependent contact angle and the drying and wetting morphological transition are studied with respect to the volume change for a spherical cap-shaped droplet placed on a spherical substrate.   The line-tension effect is included using the rigorous formula for the Helmholtz free energy in the droplet capillary model.  A morphological drying transition from a cap-shaped to a spherical droplet occurs when the substrate is hydrophobic and the droplet volume is small, similar to the transition predicted on a flat substrate.  In addition,  a morphological wetting transition from a cap-shaped to a wrapped spherical droplet occurs for a hydrophilic substrate and a large droplet volume.  The contact angle depends on the droplet size: it decreases as the droplet volume increases when the line tension is positive, whereas it increases when the line tension is negative.  The spherical droplets and wrapped droplets are stable when the line tension is positive and large.   

\end{abstract}

\maketitle

\section{\label{sec:sec1}Introduction}
The adsorption of a micro- or nano-sized liquid droplet on a structured substrate is an interesting problem.  This type of adsorption is the basis of various micro- and nano-devices based on droplets and bubbles~\cite{Seemann2012,Lohse2015}.   In particular, the understanding of liquid droplet wetting and drying onto biological structures could provide designs for new biomimetic materials~\cite{Nosonovsky2007,Song2014}.  When a liquid droplet wets a substrate to form a cap-shaped droplet, the line tension~\cite{Gibbs1906,deGennes1985,Bonn2009,Weijs2011} at the three-phase contact line should be important for determining the morphology of the droplet.   Widom~\cite{Widom1995} predicted the line-tension-induced drying transition of a flat substrate.  However, the line tension becomes important only for nanoscale droplets, because the magnitude of the line tension is quite small~\cite{Drelich1996,Pompe2000,Wang2001,Checco2003,Schimmele2007,Bonn2009}.

\begin{figure}[htbp]
\begin{center}
\includegraphics[width=0.70\linewidth]{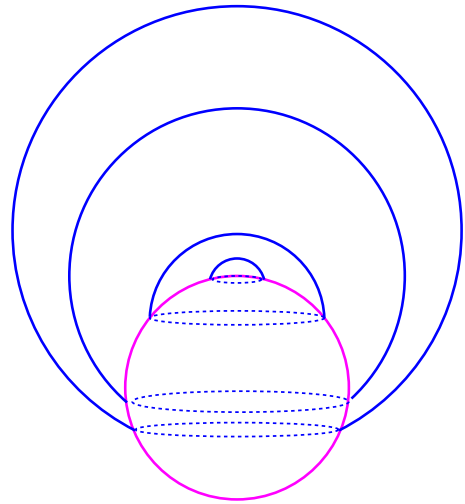}
\caption{
A cap-shaped droplet with contact angle $\theta$ and radius $r$ on a spherical substrate with radius $R$.  The droplet volume is increased by injecting the liquid through a capillary tube.
   }
\label{fig:S1}
\end{center}
\end{figure}

Unfortunately, there are few previous studies of the line-tension effect, and those that exist are mainly on flat substrates~\cite{Navascues1981,Widom1995,Blecua2006,Singha2015},  
or on {\it convex} spherical substrates~\cite{Guzzardi2007,Hienola2007,Qiu2015,Maheshwari2016}. 
We have previously derived an analytical formula for the Gibbs free energy~\cite{Iwamatsu2015a,Iwamatsu2015b} and the Helmholtz free energy~\cite{Iwamatsu2016a,Iwamatsu2016b} of a droplet placed on a spherical substrate and on the inner wall of a spherical cavity.  In this paper, we will consider the morphological phase transition induced by the volume change of a spherical droplet due to the presence of line tension, as shown in Fig.~\ref{fig:S1}.

\begin{figure}[htbp]
\begin{center}
\includegraphics[width=0.70\linewidth]{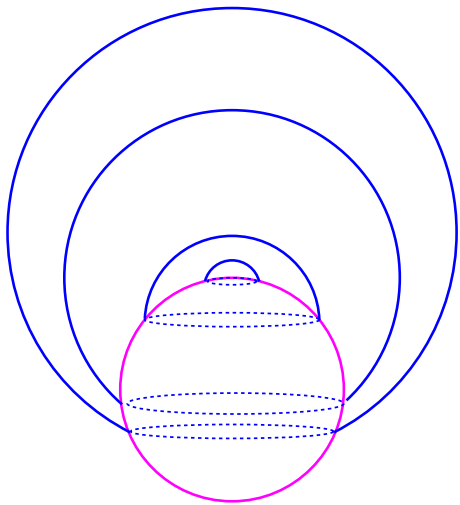}
\caption{
A cap-shaped droplets with a constant contact angle $\theta_{\rm Y}=40^{\circ}$ with different volumes (radius $r$) on a spherical substrate of radius $R$.  The droplets are always attached to the substrate and the contact lines are always on the surface of the spherical substrate.
   }
\label{fig:S2}
\end{center}
\end{figure}

When the line tension can be neglected, the droplet morphology remains cap-shaped and changes the droplet volume such that the contact angle always remains at the Young's contact angle $\theta_{\rm Y}$ (Fig.~\ref{fig:S2}).  This contact angle behavior was observed in a macroscopic-sized droplet~\cite{Tao2011}.  When the line tension effect is included, we may consider a size-dependent contact angle and a morphological transition.  Although only the drying transition~\cite{Widom1995} is possible for a positive line tension on a flat substrate, both the drying and wetting transitions are possible on a spherical substrate~\cite{Iwamatsu2016b}. 

More specifically, the droplet will form a spherical droplet on a spherical substrate in the drying transition and it will spread over the spherical substrate surface to form a wrapped sphere in the wetting transition.   In section II, we will briefly summarize the necessary formula for the Helmholtz free energy derived in our previous studies~\cite{Iwamatsu2016b}.  In section III, we will discuss the scenario for the morphological change and size-dependent contact angle induced by the changing volume using the mathematically rigorous formulation of section II.  Section IV contains the conclusion.

\section{\label{sec:sec2}Line-tension and the Helmholtz free energy of a droplet on a sphere}

We previously considered the line-tension effects on the morphology of a volatile~\cite{Iwamatsu2015a,Iwamatsu2015b} and a non-volatile~\cite{Iwamatsu2016a,Iwamatsu2016b} liquid droplet placed on a convex spherical substrates and a concave spherical cavity.   Here, we extend our previous work~\cite{Iwamatsu2016b} and focus on the evolution of the droplet morphology with respect to the liquid volume.  We consider a cap-shaped droplet of a non-volatile liquid with radius $r$ and contact angle $\theta$, placed on a convex spherical substrate of radius $R$, as shown in Fig.~\ref{fig:S1}.  For each fixed droplet volume  $V$ or radius $r$, we consider the equilibrium morphology of the droplet which corresponds to the global minimum of the Helmholtz free energy. We can then study the evolution of the droplet morphology when the liquid volume is altered.  We use the so-called {\it capillary model}, where the structure and width of the interfaces are neglected and the liquid-vapor, liquid-solid, and solid-vapor interactions are accounted for by the curvature-independent surface tensions. We consider a droplet whose radius $r$ is smaller than the capillary length so that gravity is negligible~\cite{deGennes1985,Bonn2009}.  We summarize only the necessary formulas here because all equations were previously derived~\cite{Iwamatsu2016b}.  

The Helmholtz free energy $F$ of the droplet is given by 
\begin{equation}
F=\sigma_{\rm lv}A_{\rm lv}+\Delta\sigma A_{\rm sl}+\tau L,
\label{eq:S1}
\end{equation}
and
\begin{equation}
\Delta\sigma = \sigma_{\rm sl}-\sigma_{\rm sv},
\label{eq:S2}
\end{equation}
where $A_{\rm lv}$ and $A_{\rm sl}$ are the surface areas of the liquid-vapor and solid (substrate)-liquid interfaces, respectively, and $\sigma_{\rm lv}$ and $\sigma_{\rm sl}$ are the respective surface tensions. In Eq.~(\ref{eq:S2}), $\Delta\sigma$ represents the free energy gain as the solid-vapor interface with the surface tension $\sigma_{\rm sv}$ is replaced by the solid-liquid interface with surface tension $\sigma_{\rm sl}$ when a liquid droplet covers the substrate. The effect of the line tension $\tau$ is given by the last term of Eq.~(\ref{eq:S1}), where $L$ denotes the length of the three-phase contact line.

The Helmholtz free energy of a droplet with contact angle $\theta$ and radius $r$ is derived from Eq.~(\ref{eq:S1}), and is given by~\cite{Iwamatsu2016b}
\begin{equation}
F=4\pi R^2\sigma_{\rm lv}f\left(\rho,\theta\right),
\label{eq:S3}
\end{equation}
with
\begin{equation}
f\left(\rho,\theta\right)=\rho\frac{\left(\rho+\zeta\right)^2-1}{4\zeta}-\cos\theta_{\rm Y}\frac{\rho^{2}-\left(1-\zeta\right)^{2}}{4\zeta}+\tilde{\tau}\frac{\rho\sin\theta}{2\zeta},
\label{eq:S4}
\end{equation}
where
\begin{equation}
\zeta=\sqrt{1+\rho^{2}-2\rho\cos\theta}
\label{eq:S5}
\end{equation}
and
\begin{equation}
\rho=\frac{r}{R}
\label{eq:S6}
\end{equation}
is the size parameter of the droplet.  The Young's contact angle $\theta_{\rm Y}$ is defined by the classical Young's equation~\cite{Young1805} on a flat substrate, 
\begin{equation}
\Delta\sigma+\sigma_{\rm lv}\cos\theta_{\rm Y}=0,
\label{eq:S7}
\end{equation}
and the scaled line tension $\tilde{\tau}$ is defined by
\begin{equation}
\tilde{\tau}=\frac{\tau}{\sigma_{\rm lv}R}.
\label{eq:S8}
\end{equation}

The equilibrium contact angle $\theta_{\rm e}$ is determined by extremizing the free energy in Eq.~(\ref{eq:S3}) under the constant volume subsidiary condition given by
\begin{equation}
V=\frac{4\pi}{3}R^{3}\omega\left(\rho,\theta\right)
\label{eq:S9}
\end{equation}
with
\begin{eqnarray}
\omega\left(\rho,\theta\right)&=&\frac{1}{16\zeta}\left(\zeta-1+\rho\right)^{2} \nonumber \\
&&\times\left[3\left(1+\rho\right)^{2}-2\zeta\left(1-\rho\right)-\zeta^{2}\right],
\label{eq:S10}
\end{eqnarray}
which leads to the equation that determines $\theta_{\rm e}$ or radius $\rho_{\rm e}$ written as
\begin{equation}
\cos\theta_{\rm Y}-\cos\theta_{\rm e}-\tilde{\tau}\frac{1-\rho_{\rm e}\cos\theta_{\rm e}}{\rho_{\rm e}\sin\theta_{\rm e}}=0.
\label{eq:S11}
\end{equation}
Note that the droplet radius $\rho_{\rm e}$ will be determined from the equilibrium contact angle $\theta_{\rm e}$ when the droplet volume $V$ is fixed.  This equation, known as the generalized Young's equation, reduces to Eq.~(\ref{eq:S7}) when $\tilde{\tau}=0$.  In this case, both angles $\theta_{\rm e}$ and $\theta_{\rm Y}$ are identical.

It may also be noted from Eq.~(\ref{eq:S11}) that the line tension does not affect the value of the equilibrium contact angle $\theta_{\rm e}$ when this attains the value of a characteristic contact angle $\theta_{\rm c}$ defined by 
\begin{equation}
1-\rho_{\rm c}\cos\theta_{\rm c}=0,
\label{eq:S12}
\end{equation}
where $\rho_{\rm c}$ is the radius of the droplet when the contact angle is $\theta_{\rm c}$.  In this case, the equilibrium contact angle is given simply by the Young's contact angle $\theta_{\rm Y}$, and the contact line coincides with the equator of the spherical substrate~\cite{Iwamatsu2015b}.  In other words, the equilibrium contact angle is simply given by the Young's contact angle ($\theta_{\rm e}=\theta_{\rm Y}$) and it is not affected by the presence of the line tension when the contact line coincides with the equator  ($\theta_{\rm e}=\theta_{\rm c}$).  We can determine the value of the intrinsic Young's contact angle $\theta_{\rm Y}$ without the line-tension effect by simply measuring the contact angle when the contact line coincides with the equator of a spherical substrate. 

To specify the droplet volume, we characterize the droplet volume by the size parameter $\rho_{180}=\rho\left(\theta=180^{\circ}\right)$ when it is a complete sphere residing on a spherical substrate\cite{Iwamatsu2016b}. Then, we have
\begin{equation}
\omega\left(\rho,\theta\right)=\omega\left(\rho=\rho_{180},\theta=180^{\circ}\right)=\rho_{180}^{3},
\label{eq:S13}
\end{equation}
which will determine the size parameter $\rho=\rho\left(\theta\right)$ as a function of the contact angle $\theta$.  Then, the Helmholtz free energy Eq.~(\ref{eq:S4}) of the droplet with a fixed volume $V$ or a size $\rho_{180}$ becomes a function of the contact angle $\theta$.  

The equilibrium contact angle $\theta_{\rm e}$ determined from Eq.~(\ref{eq:S11}) is the local extremum of the Helmholtz free energy Eq.~(\ref{eq:S4}) of the cap-shaped droplet given by
\begin{equation}
F_{\rm cap}=4\pi R^{2}\sigma_{\rm lv}f_{\rm e}
\label{eq:S14}
\end{equation}
with
\begin{eqnarray}
f_{\rm e} &=& \frac{\left(-1+\rho_{\rm e}+\zeta_{\rm e}\right)^{2}\left(\cos\theta_{\rm e}+1+\zeta_{\rm e}\right)}{4\zeta_{\rm e}}
\nonumber \\
&&+\tilde{\tau}\frac{\left(-1+\rho_{\rm e}\cos\theta_{\rm e}+\zeta_{\rm e}\right)}{2\rho_{\rm e}\sin\theta_{\rm e}},
\label{eq:S15}
\end{eqnarray}
where $\zeta_{\rm e}$ and $\rho_{\rm e}$ are Eq.~(\ref{eq:S5}) and Eq.~(\ref{eq:S6}), respectively, when $\theta=\theta_{\rm e}$.

In addition to these cap-shaped droplets, we may consider a spherical droplet residing on  a spherical substrate~\cite{Iwamatsu2016b}, whose free energy is given by Eq.~(\ref{eq:S3}) when $\theta=180^{\circ}$: 
\begin{equation}
F_{\rm sphere}=4\pi R^{2}\sigma_{\rm lv}f_{180}
\label{eq:S16}
\end{equation}
where
\begin{equation}
f_{180}=\rho_{180}^{2}.
\label{eq:S17}
\end{equation}
Furthermore,  we may consider a droplet to completely wrap the spherical substrate~\cite{Kuni1996,Bieker1998,Bykov2006} when $\rho>1$ and $\theta=0$.  The free energy~\cite{Iwamatsu2016b} is given again by Eq.~(\ref{eq:S3}) when $\theta=0^{\circ}$: 
\begin{equation}
F_{\rm wrap}=4\pi R^{2}\sigma_{\rm lv}f_{0},
\label{eq:S18}
\end{equation}
where
\begin{equation}
f_{0}=\rho_{0}^{2}-\cos\theta_{\rm Y}
\label{eq:S19}
\end{equation}
and $\rho_{0}$ is the size parameter when the contact angle is $\theta=0^{\circ}$, which is related to the size $\rho_{180}$ through the conservation of the droplet volume given by
\begin{equation}
\rho_{0}^3-1=\rho_{180}^{3}.
\label{eq:S20}
\end{equation}

The transformation between a cap-shaped and a detached, spherical morphology is realized when $F_{\rm sphere}=F_{\rm cap}$.  We refer to this transformation as the {\it drying transition}, though the drying transition of the surface-phase transition is used for infinite open systems~\cite{Dietrich1988,Bonn2009} induced by the change of the chemical potential or the vapor pressure. In this context, the spherical droplet corresponds to the drying state and the cap-shaped droplet corresponds to the incomplete wetting state.  Similarly, we refer to the morphological transformation between the cap-shaped to wrapped spherical droplet, which is realized when $F_{\rm wrap}=F_{\rm cap}$, as the {\it wetting transition}.  Therefore, the wrapped spherical droplet corresponds to the wetting state.
We analyze the volume-induced transformation of a droplet placed on a spherical substrate in the next section.

\section{\label{sec:sec3}   Morphological transition of a droplet induced by the volume change}

\begin{figure}[htbp]
\begin{center}
\centering
\includegraphics[width=0.90\linewidth]{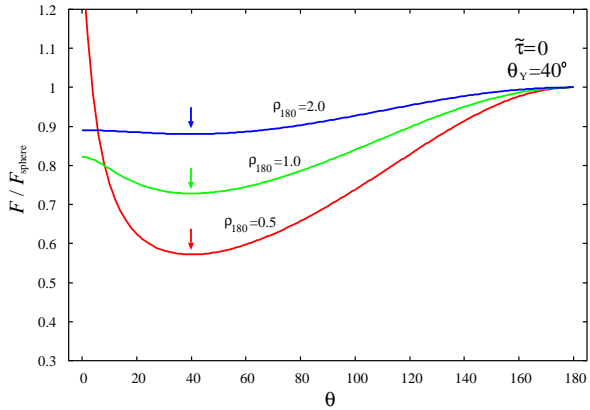}
\caption{
The Helmholtz free energy landscape defined by Eq.~(\ref{eq:S4}) when $\theta_{\rm Y}=40^{\circ}$ and $\tilde{\tau}=0$. The global minimum is always fixed at the intrinsic Young's contact angle $\theta_{\rm Y}$ indicated by down arrows, as predicted by the generalized Young's equation (\ref{eq:S11}) with $\tilde{\tau}=0$.
   }
\label{fig:S3}
\end{center}
\end{figure}

Visualization of the free-energy landscape of Eq.~(\ref{eq:S4}) is the easiest way to determine the stable morphology of a droplet placed on a spherical substrate. Figure~\ref{fig:S3} shows the free-energy landscape of a droplet with a different volume or radius $\rho_{180}$ when the line tension is not included ($\tilde{\tau}=0$).  The absolute minimum of the free energy is always at $\theta_{\rm Y}=40^{\circ}$ (Fig.~\ref{fig:S2}) because the equilibrium contact angle is determined from Eq.~(\ref{eq:S11}) with $\tilde{\tau}=0$, which shows that the equilibrium contact angle $\theta_{\rm e}$ is simply given by the intrinsic Young's contact angle $\theta_{\rm e}=\theta_{\rm Y}$.

\begin{figure}[htbp]
\begin{center}
\includegraphics[width=0.90\linewidth]{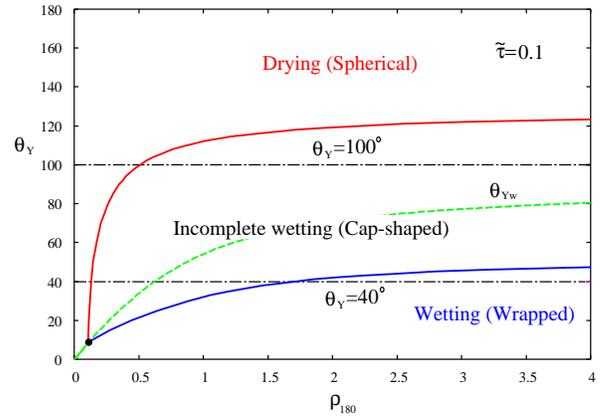}
\caption{
The morphological phase diagram when $\tilde{\tau}=0.1$.  The Young's contact angle for the drying transition (upper curve) and that for the wetting transition (lower curve) are shown.  The middle dash curve represents the wetting-drying boundary $\theta_{\rm Y,w}$ defined by Eq.~(\ref{eq:S21}).  The cap-shaped droplet is stable in the region between the upper and lower curve.  These three curves merge at the triple point indicated by a filled circle, where  cap-shaped, spherical and wrapped spherical morphologies coexist.  Two horizontal dot-dash lines represent the two routes with Young's contact angle $\theta_{\rm Y}=40^{\circ}$ and $100^{\circ}$.  By increasing the droplet radius $\rho_{180}$ along these two lines, both the drying and the wetting transitions can occur when $\theta_{\rm Y}=40^{\circ}$.  However, only the drying transition can occur when $\theta_{\rm Y}=100^{\circ}$. 
   }
\label{fig:S4}
\end{center}
\end{figure}

When the line tension is positive, we may consider a wetting transition between a cap-shaped droplet and a wrapped spherical droplet and a drying transition between a cap-shaped droplet and a spherical droplet~\cite{Iwamatsu2016b}.  Figure \ref{fig:S4} shows the morphological phase diagram when $\tilde{\tau}=0.1$.  The upper curve represents the drying transition line, above which the equilibrium structure with the lowest free energy is a spherical droplet with the contact angle $\theta=180^{\circ}$.  Then,  the spherical substrate is in the drying state.   The lower curve represents the wetting transition line, below which the equilibrium structure with the lowest free energy is a wrapped spherical droplet with the contact angle $\theta=0^{\circ}$.  These two curves merge at the triple point represented by the filled circle on the middle curve in Fig.~\ref{fig:S4}, where the three morphologies, cap-shaped, spherical and wrapped-spherical coexist. This middle curve represent the wetting-drying boundary $\theta_{\rm Y,w}$ determined from $F_{\rm sphere}=F_{\rm wrap}$ in Eqs.~(\ref{eq:S16}) and (\ref{eq:S18}), which leads to~\cite{Iwamatsu2016a}
\begin{equation}
\theta_{\rm Y,w}=\cos^{-1}\left[\left(1+\rho_{180}^{3}\right)^{2/3}-\rho_{180}^{2}\right],
\label{eq:S21}
\end{equation}
where the conservation of the droplet volume given by Eq.~(\ref{eq:S20}) was used.  When the intrinsic Young's contact angle is larger than {the wetting-drying boundary given by }Eq.~(\ref{eq:S21}) ($\theta_{\rm Y}>\theta_{\rm Y,w}$), a spherical droplet is relatively more stable than a wrapped droplet.  However, the cap-shaped droplet with incomplete wetting is more stable than the spherical droplet and the wrapped spherical droplet within the region between the upper curve and the lower curve. This cap-shaped droplet is metastable, whereas only the spherical and wrapped spherical droplet are stable, when the radius $\rho_{180}$ is smaller than the triple point (Fig.~\ref{fig:S4}).

The wettability of a substrate is characterized by the Young's contact angle $\theta_{\rm Y}$ (e.g., $\theta_{\rm Y}=40^{\circ}$ in Fig.~\ref{fig:S4}).  The liquid droplet may undergo a morphological drying transition from a spherical droplet with the equilibrium contact angle $\theta_{\rm e}=180^{\circ}$ to a cap-shaped droplet with $\theta_{\rm e}\simeq 80^{\circ}$ at $\rho_{180}\simeq 0.131$ (the intersection of the upper curve and the horizontal curve $\theta_{\rm Y}=40^{\circ}$ in Fig.~\ref{fig:S4}) by increasing the droplet volume.  In addition, the droplet may undergo a morphological wetting transition from a cap-shaped droplet to a completely wrapped spherical droplet with the equilibrium contact angle $\theta_{\rm e}=0^{\circ}$ at $\rho_{180}\simeq 1.70$  (the intersection of the lower curve and the horizontal curve $\theta_{\rm Y}=40^{\circ}$ in Fig.~\ref{fig:S4}).  The wetting transition would be suppressed for a more hydrophobic substrate ($\theta_{\rm Y}=100^{\circ}$).  Only the drying state of a spherical droplet can appear at $\rho_{180}\simeq 0.510$ (Fig.~\ref{fig:S4}).

\begin{figure}[htbp]
\begin{center}
\subfigure[]
{
\includegraphics[width=0.9\linewidth]{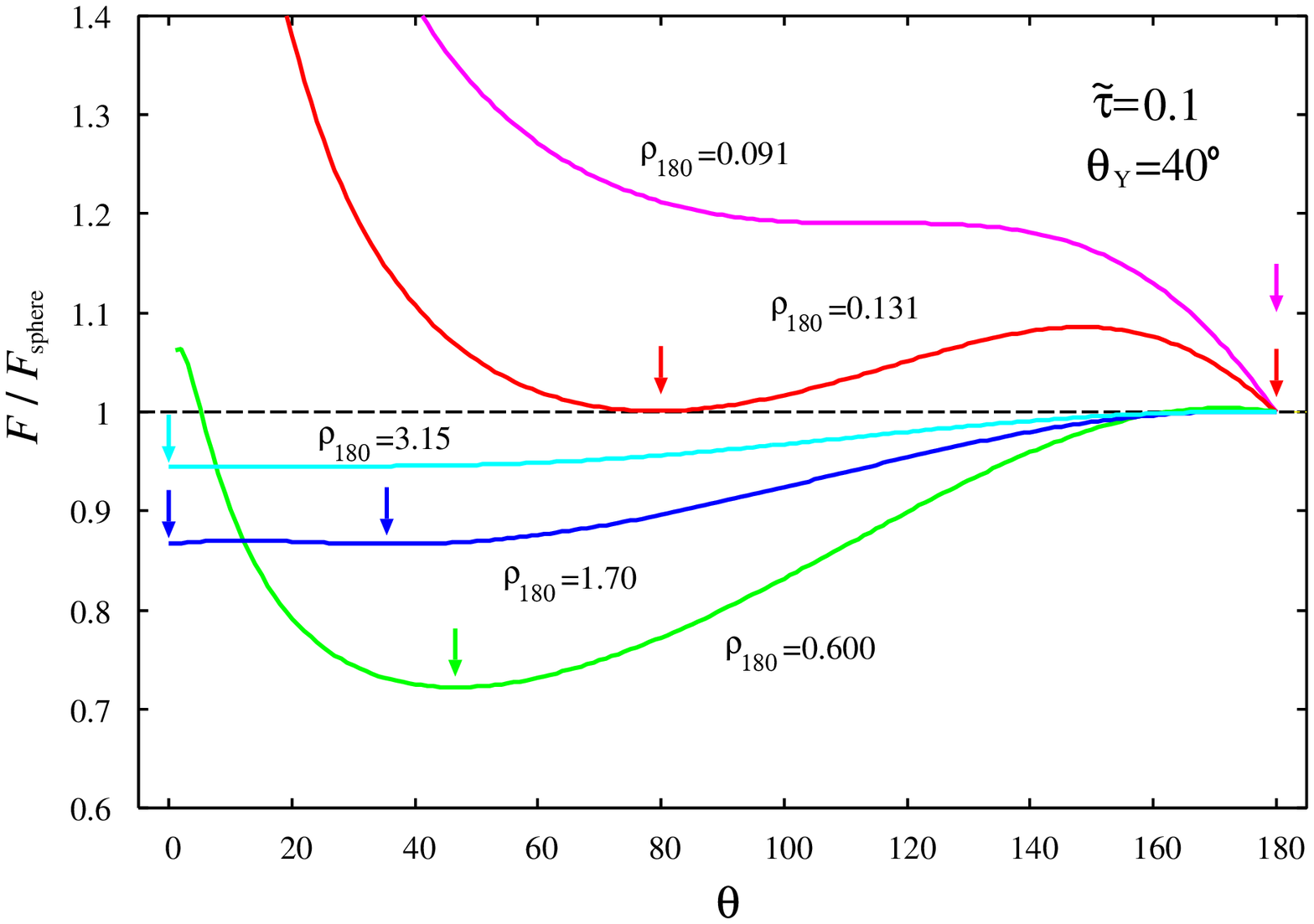}
\label{fig:S5a}
}
\subfigure[]
{
\includegraphics[width=0.9\linewidth]{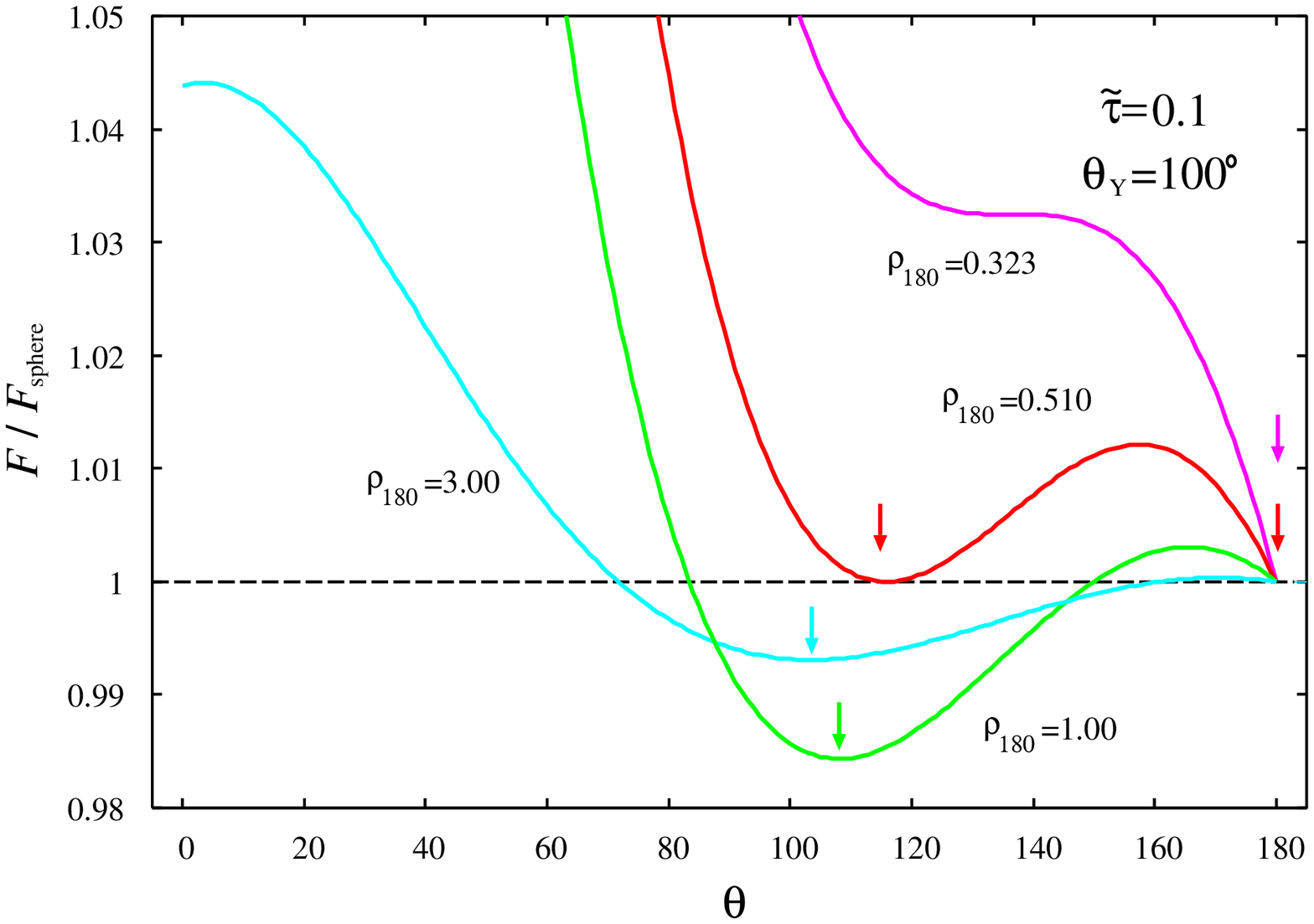}
\label{fig:S5b}
}
\end{center}
\caption{
(a) The Helmholtz free energy landscape defined by Eq.~(\ref{eq:S4}) when $\theta_{\rm Y}=40^{\circ}$ and $\tilde{\tau}=0.1$. There are three types of global minima indicated by down arrows which correspond to a spherical droplet ($\theta_{\rm e}=180^{\circ}$), a wrapped spherical droplet  ($\theta_{\rm e}=0^{\circ}$) and a cap-shaped droplet.  A cap-shaped droplet lose its stability at $\rho_{180}=0.091$ and $3.15$.  At the drying transition ($\rho_{180}=0.131$) and the wetting transition ($\rho_{180}=1.70$), the two minima have the same free energy and the corresponding two morphologies can coexist.  (b) The same as (a) when $\theta_{\rm Y}=100^{\circ}$ and $\tilde{\tau}=0.1$.  In this hydrophobic case, only the drying transition can occur at $\rho_{180}=0.510$ and the metastable cap-shaped droplet becomes unstable at $\rho_{180}=0.323$. 
 } 
\label{fig:S5}
\end{figure}

Figure \ref{fig:S5} shows the free energy landscape when $\tilde{\tau}=0.1$ and (a) $\theta_{\rm Y}=40^{\circ}$ and (b) $\theta_{\rm Y}=100^{\circ}$.  When $\theta_{\rm Y}=40^{\circ}$, not only the drying transition but also the wetting transition is possible from Fig.~\ref{fig:S4}.  In fact, the drying transition occurs at $\rho_{180}\simeq0.131$, where the two global minima at $\theta=180^{\circ}$ and $\theta\simeq 80^{\circ}$ coexist (Fig.~\ref{fig:S5}(a)).  The wetting transition occurs at $\rho_{180}\simeq1.70$, where the two global minima at  $\theta=0^{\circ}$ and  $\theta\simeq 35^{\circ}$ coexist.   However, the free energy barrier exists between the incomplete-wetting cap-shaped droplet and the drying spherical droplet.  Also, the free energy barrier exists between the incomplete-wetting cap-shaped droplet and wetting wrapped droplet.  Hence, we require extra work to induce these transitions at the same volume.  Therefore, the morphological drying transition and the wetting transition are similar to the first-order phase transition. Beyond these two transition points, the cap-shaped droplet can exist as a metastable droplet up to the stability limit $\rho_{180}\simeq0.091$ and  $\rho_{180}\simeq3.15$ (Fig.~\ref{fig:S5}(a)), which are similar to the spinodal of the first-order phase transition, where a cap-shaped droplet with a finite contact angle becomes unstable. The cap-shaped droplet will transform into a spherical droplet or a wrapped spherical droplet spontaneously at these spinodals without crossing the free-energy barrier.

On the other hand, only the drying transition is possible when $\theta_{\rm Y}=100^{\circ}$ as can be seen from Fig.~\ref{fig:S4}.  In fact, only the drying transition is observed in Fig.~\ref{fig:S5}(b) where the drying transition occurs at $\rho_{180}\simeq0.510$.  The cap-shaped droplet is always stable when $\rho_{180}>0.510$.  The stability limit of the cap-shaped droplet occurs at $\rho_{180}\simeq0.323$ (Fig.~\ref{fig:S5}(b)).

\begin{figure}[htbp]
\begin{center}
\includegraphics[width=0.90\linewidth]{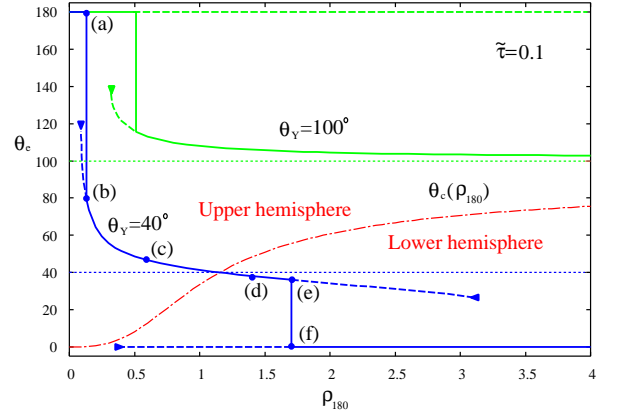}
\caption{
The equilibrium contact angle $\theta_{\rm e}$ (solid curve) as a function of the droplet size $\rho_{180}$ when the substrate is hydrophilic ($\theta_{\rm Y}=40^{\circ}$) and hydrophobic ($\theta_{\rm Y}=100^{\circ}$) for a positive line tension $\tilde{\tau}=0.1$.  The long dash curves represent metastable droplets and the filled triangles indicate the stability limits of the metastable states.  The short dash curves represent the intrinsic Young's contact angle $\theta_{\rm Y}=40^{\circ}$ and $100^{\circ}$.  The middle dot-dash curve represents the characteristic contact angle $\theta_{\rm c}$ defined by Eq.~(\ref{eq:S12}) as a function of the droplet size $\rho_{180}$. The contact line is located on the upper hemisphere when $\theta_{\rm e}$ is above this curve ($\theta_{\rm e}>\theta_{\rm c}$).  The metastable droplet with $\theta_{\rm e}=180^{\circ}$ for $\theta_{\rm Y}=40^{\circ}$ is not shown since it overlaps that for $\theta_{\rm Y}=100^{\circ}$. The morphologies, which corresponds to the filled circles labeled by (a)-(f) will be shown in Fig.~\ref{fig:S7}. 
   }
\label{fig:S6}
\end{center}
\end{figure}

Figure~\ref{fig:S6} shows the size dependence of the contact angle when $\theta_{\rm Y}=40^{\circ}$ and $\theta_{\rm Y}=100^{\circ}$.  The equilibrium contact angle decreases monotonically and approaches the intrinsic Young's contact angle $\theta_{\rm Y}$ from above for a hydrophobic substrate ($\theta_{\rm Y}=100^{\circ}$).  The same behavior for the size-dependent contact angle was considered on a flat substrate~\cite{Gaydos1987}, which has been frequently observed~\cite{Wang2001}.  However, on a spherical substrate there is a drying transition where the contact angle jumps from $\theta_{\rm e}=180^{\circ}$ to $\theta_{\rm e}\sim \theta_{\rm Y}$ when  $\rho_{180}$ is small.  

In contrast, the contact angle becomes $\theta_{\rm e}=0^{\circ}$ and a wetting transition of a wrapped droplet appears for a large droplet ($\rho_{180}\gtrsim 1.7$) in addition to the drying transition of a spherical droplet when the substrate is hydrophilic ($\theta_{\rm Y}=40^{\circ}$).  The equilibrium contact angle decreases monotonically~\cite{Gaydos1987}. We also show $\theta_{\rm c}$ (defined by Eq.~(\ref{eq:S12})) as a function of the droplet size $\rho_{180}$. Since, the contact line coincides with the equator of the substrate when $\theta_{\rm e}=\theta_{\rm c}$, the contact line is located on the upper hemisphere above this curve and on the lower hemisphere below this curve.  Therefore, the contact line crosses the equator from the upper hemisphere to the lower hemisphere at $\rho_{180}\sim 1.2$ as  $\rho_{180}$ increases when $\theta_{\rm Y}=40^{\circ}$. The equilibrium contact angle at the equator is given by the intrinsic Young's contact angle $\theta_{\rm e}=\theta_{\rm Y}=40^{\circ}$. Since $\theta_{\rm c}\rightarrow 90^{\circ}$ as $\rho_{180}\rightarrow\rho_{\rm c}\rightarrow\infty$, the contact line of a droplet eventually crosses the equator on the hydrophilic substrate with $\theta_{\rm Y}<90^{\circ}$ by increasing the droplet volume.  Then, $\theta_{\rm Y}$ could be measured directly when the three-phase contact line crosses the equator of the spherical substrate.

The long dash curves in Fig.~\ref{fig:S6} represent the metastable droplets and the filled triangle indicates the stability limits of the metastable states similar to the spinodal of the first-order phase transitions.  The metastable spherical droplet with $\theta_{\rm e}=180^{\circ}$ could persist at a volume larger than $\rho_{180}=4.0$.  In contrast, the metastable wrapped spherical droplet with $\theta_{\rm e}=0^{\circ}$ and the metastable cap-shaped droplet have stability limits.  Therefore, we would observe hysteresis when $\rho_{180}$ increases and decreases as in the case of the first order phase transitions.  

\begin{figure}[htbp]
\begin{center}
\includegraphics[width=0.90\linewidth]{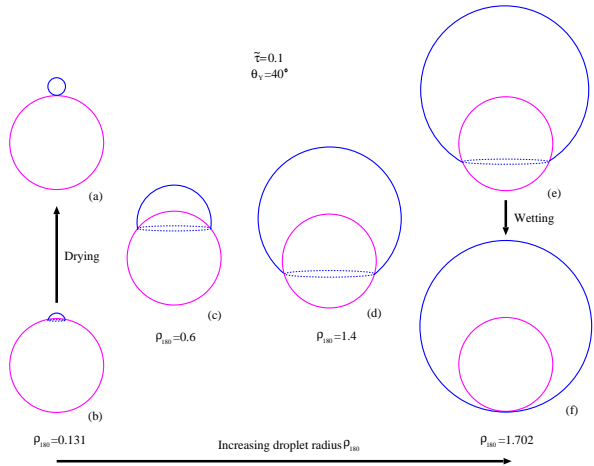}
\caption{
The evolution of a droplet morphology, which corresponds to the states labelled (a)-(f) in Fig.~\ref{fig:S6}, induced by the volume change of the droplet on a hydrophilic surface when $\theta_{\rm Y}=40^{\circ}$ and $\tilde{\tau}=0.1$.  In contrast to Fig.~\ref{fig:S1}, a drying spherical droplet and a wetting wrapped spherical droplet appear on a spherical substrate when the line tension is finite and positive.
   }
\label{fig:S7}
\end{center}
\end{figure}

Based on the scenario that has emerged from the above analysis, we show pictorially in Fig.~\ref{fig:S7} the morphological transition of a droplet place on a spherical substrate, when $\theta_{\rm Y}=40^{\circ}$ and $\tilde{\tau}=0.1$, where the morphologies of the droplet that correspond to the states labeled (a)-(f) in Fig.~\ref{fig:S6} are shown.  Initially, a small volume droplet would be spherical when residing on the substrate.  The droplet would spread on the substrate and transform from a spherical  (Figs.~\ref{fig:S7}(a)) to a cap-shaped droplet  (Fig.~\ref{fig:S7}(b)) at a larger volume of $\rho_{180}\simeq 0.131$.  This is similar to the first-order phase transition (drying transition) with a free-energy barrier (see also Figs.~\ref{fig:S5}(a) and \ref{fig:S6}).  On further increasing the droplet volume, the contact line moves downward from the upper hemisphere (Fig.~\ref{fig:S7}(c)) to the lower hemisphere (Fig.~\ref{fig:S7}(d)).   When the volume reaches the wetting transition point of radius $\rho_{180}\simeq 1.702$, the wetting transition from the cap-shaped (Fig.~\ref{fig:S7}(e)) to a wrapped sphere (Fig.~\ref{fig:S7}(f)) occurs  (see also Figs.~\ref{fig:S5}(a) and \ref{fig:S6}).  Therefore, both the line-tension-induced drying transition~\cite{Widom1995} and the wetting transition would be observed on a spherical substrate when the line tension is positive.  Both a hydrophobic substrate (e.g., $\theta_{\rm Y}=100^{\circ}$) and a hydrophilic substrate (e.g., $\theta_{\rm Y}=40^{\circ}$) can be {\it super hydrophobic} with the contact angle $\theta=180^{\circ}$ when the droplet is small on a spherical substrate (Fig.~\ref{fig:S6}) and on a flat substrate~\cite{Widom1995}.  However, a hydrophilic substrate (e.g., $\theta_{\rm Y}=40^{\circ}$) can also be {\it super hydrophilic} on a spherical substrate with the contact angle $\theta=0^{\circ}$ (Fig.~\ref{fig:S6}) when the droplet is large.

\begin{figure}[htbp]
\begin{center}
\includegraphics[width=0.90\linewidth]{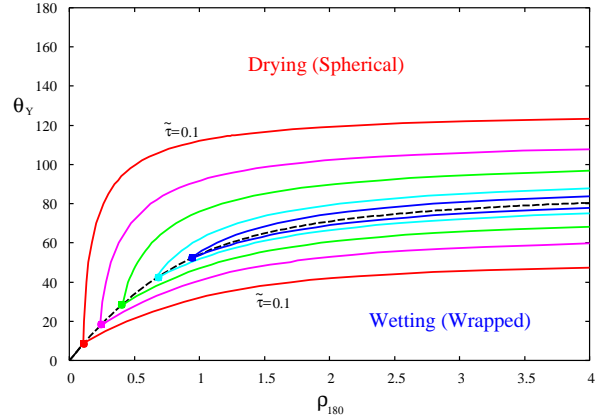}
\caption{
The morphological phase diagram similar to Fig.~\ref{fig:S4} when $\tilde{\tau}=0.1 \mbox{(outermost)}, 0.2, 0.3. 0.4$, and $0.45 \mbox{(innermost)}$. The region between the upper curve and the lower curve, which corresponds to a cap-shaped droplet, becomes narrower and the triple point moves to infinity as the magnitude of the line tension $\tilde{\tau}$ is increased towards the upper limit $\tilde{\tau}_{\rm u}=1/2$.
   }
\label{fig:S8}
\end{center}
\end{figure}

\begin{figure}[htbp]
\begin{center}
\subfigure[]
{
\includegraphics[width=0.9\linewidth]{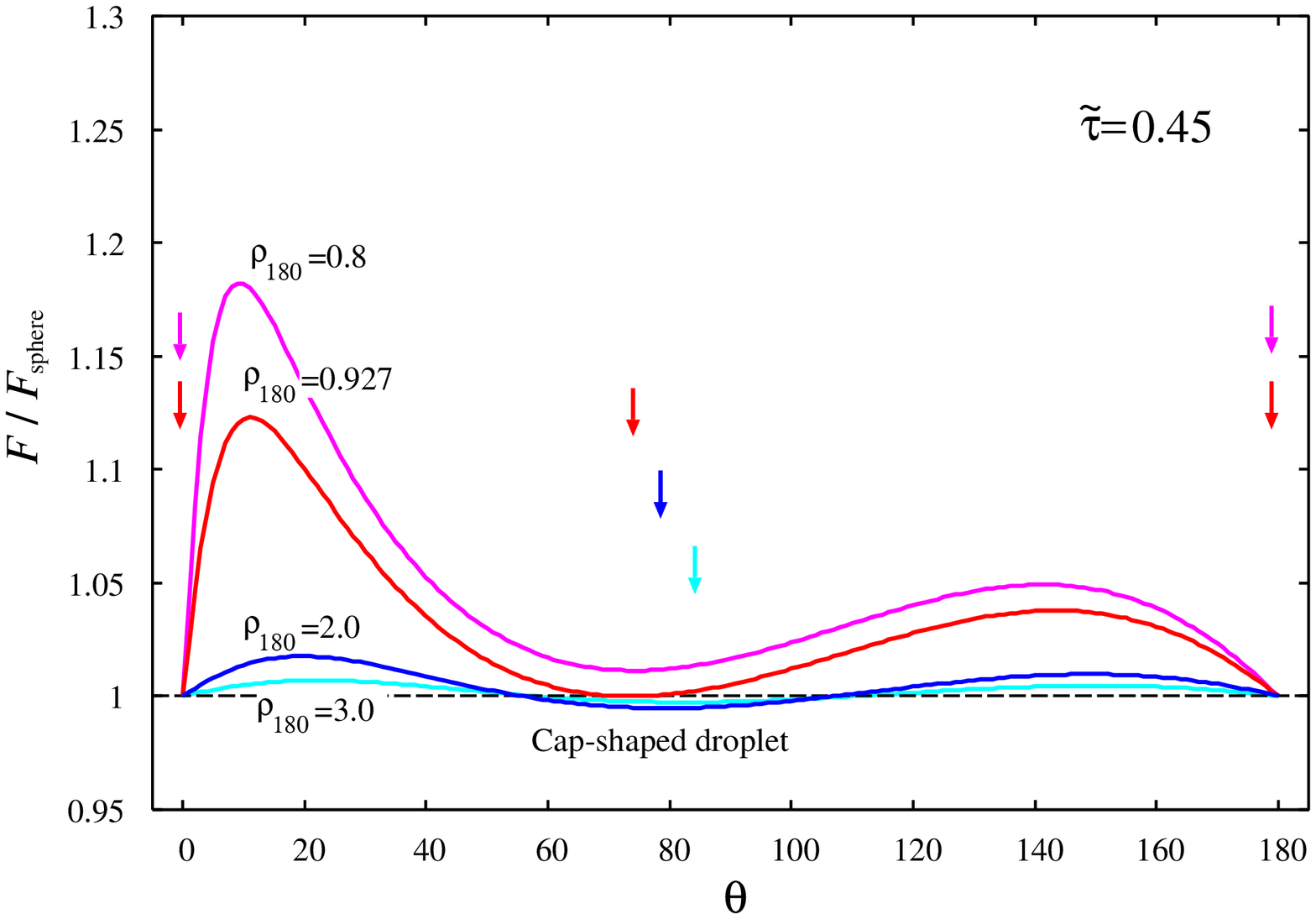}
\label{fig:S9a}
}
\subfigure[]
{
\includegraphics[width=0.9\linewidth]{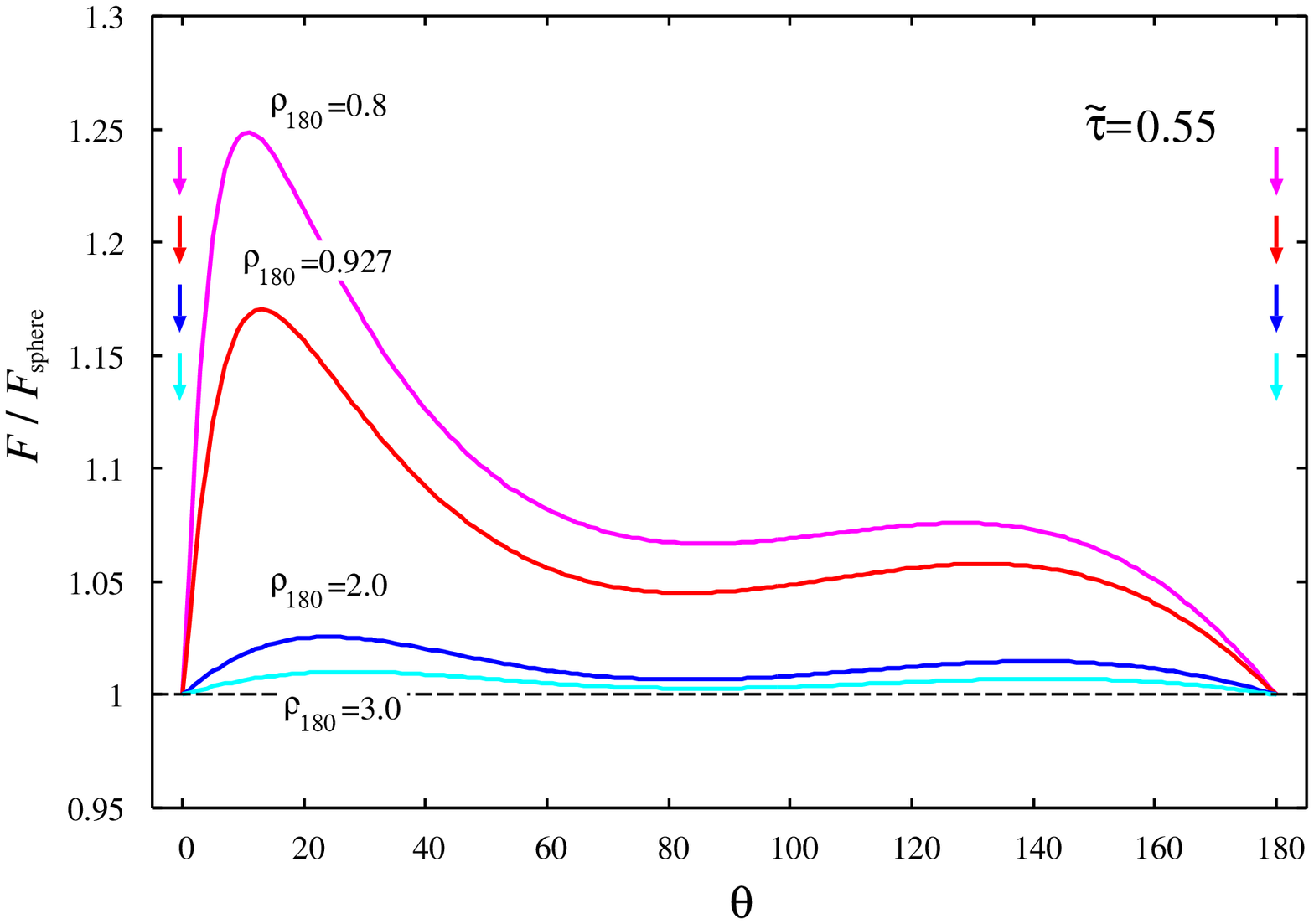}
\label{fig:S9b}
}
\end{center}
\caption{
The Helmholtz free energy landscape defined by Eq.~(\ref{eq:S4}) along the wetting-drying boundary $\theta_{\rm Y,w}$  when  (a) $\tilde{\tau}=0.45 (<\tilde{\tau}_{\rm u}=1/2)$ and (b) $\tilde{\tau}=0.55 (>\tilde{\tau}_{\rm u}=1/2)$ . The global minimum (down arrow) appears at the wetting state at $\theta=0^{\circ}$ and at the drying state at $\theta=180^{\circ}$ since the two states have the same free energy along the wetting-drying boundary $\theta_{\rm Y,w}$.  The global minimum that corresponds to the cap-shaped droplet appears only when  $\tilde{\tau}=0.45<1/2$, while the cap-shaped droplet becomes metastable and only a spherical droplet and a wrapped spherical droplet appear when $\tilde{\tau}=0.55>1/2$.  The cap-shaped, spherical, and wrapped spherical droplet can coexist at the triple point $\rho_{180}=0.927$ when $\tilde{\tau}=0.45$.
 } 
\label{fig:S9}
\end{figure}

A similar situation can occur when $\tilde{\tau}$ is larger than $0.1$.  Figure~\ref{fig:S8} shows the morphological phase diagram for $\tilde{\tau}=0.1, 0.2, 0.3, 0.4$, and $0.45$.  We observe that the triple point on the wetting-drying boundary curve $\theta_{\rm Y,w}$ moves towards the larger radius $\rho_{180}$ and the region for the cap-shaped droplet becomes narrower and is absorbed into the boundary $\theta_{\rm Y, w}$.  As the volume of the triple point tend to infinity ($\rho_{180}\rightarrow\infty$) and $\theta_{\rm Y}\rightarrow \theta_{\rm Y,w}\rightarrow 90^{\circ}$ from Eq.~(\ref{eq:S21}), the free energy of the cap-shaped droplet becomes $f_{\rm cap}\rightarrow f\left(\rho\rightarrow\rho_{180}\rightarrow\infty,\theta= 90^{\circ}\right)=\rho_{180}^{2}-1/4+\tilde{\tau}/2$ from Eq.~(\ref{eq:S4}).  Since the triple point exists only when $f_{\rm cap}<f_{180}$ from Eq.~(\ref{eq:S17}), the scaled line tension must satisfy $\tilde{\tau}<1/2$.  Therefor, on further increase of the line tension towards the upper limit $\tilde{\tau}\rightarrow \tilde{\tau}_{u}=1/2$, the triple point tends to infinity along the curve $\theta_{\rm Y}=\theta_{\rm Y,w}$ and the region for the cap-shaped droplet disappears.

We show the free-energy landscape along the wetting-drying boundary $\theta_{\rm Y,w}$ when (a) $\tilde{\tau}=0.45<1/2$ and (b) $\tilde{\tau}=0.55>1/2$ in Fig.~\ref{fig:S9}.  When $\tilde{\tau}=0.45$, the triple point exists at $\rho_{180}\simeq 0.927$.  The cap-shaped droplet can appear as it has the free energy lower than those of the spherical droplet at $\theta=180^{\circ}$ and the wrapped spherical droplet at $\theta=0^{\circ}$.  In fact, there appears a free energy minimum near $\theta_{\rm e}\sim 80^{\circ}$ in Fig.~\ref{fig:S9}(a) when $\rho_{180}=2.0$ and $3.0$.  When $\tilde{\tau}=0.55$, on the other hand, the triple point disappears and only the spherical droplet and the wrapped spherical droplet can appear.

\begin{figure}[htbp]
\begin{center}
\subfigure[]
{
\includegraphics[width=0.9\linewidth]{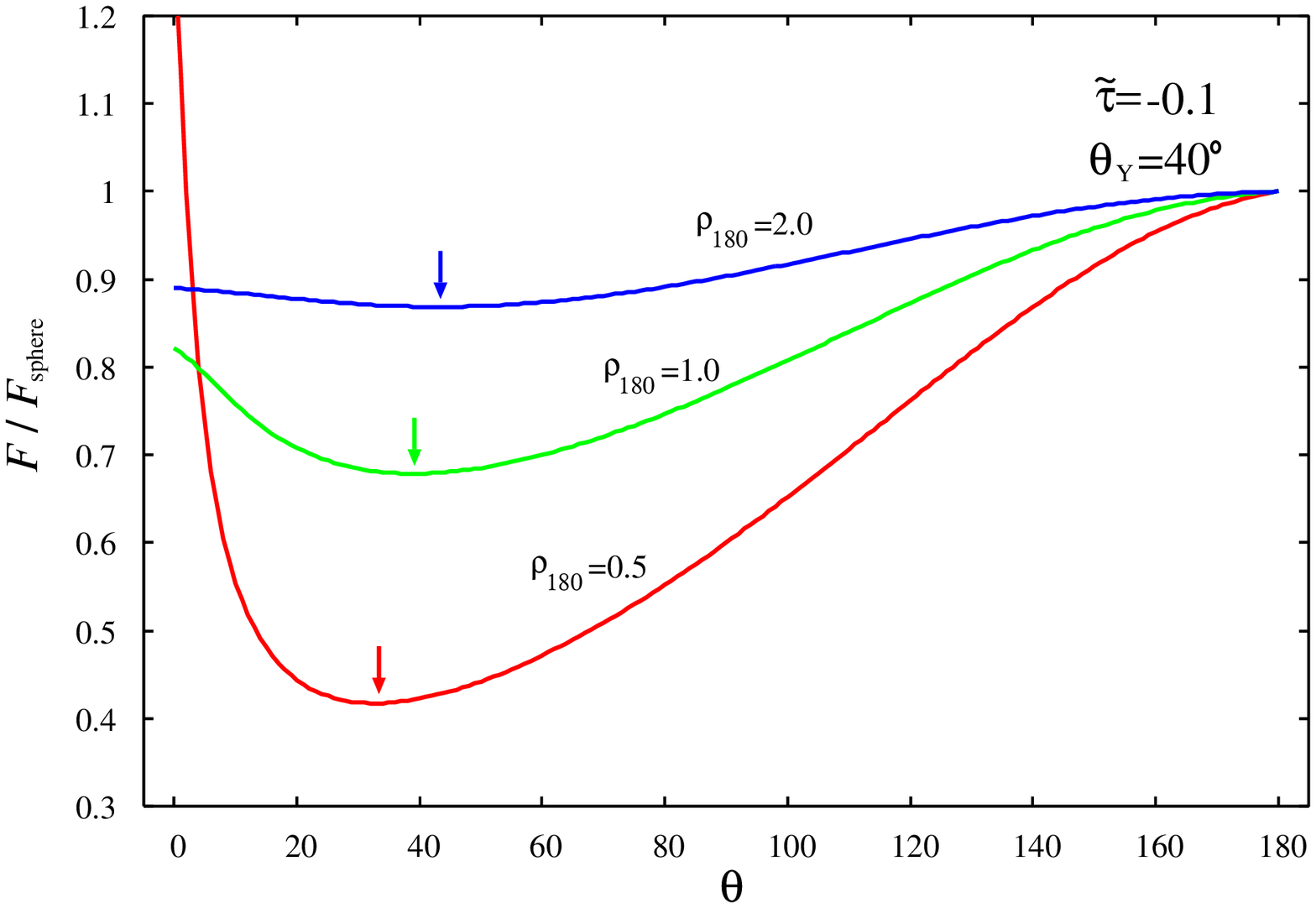}
\label{fig:S10a}
}
\subfigure[]
{
\includegraphics[width=0.9\linewidth]{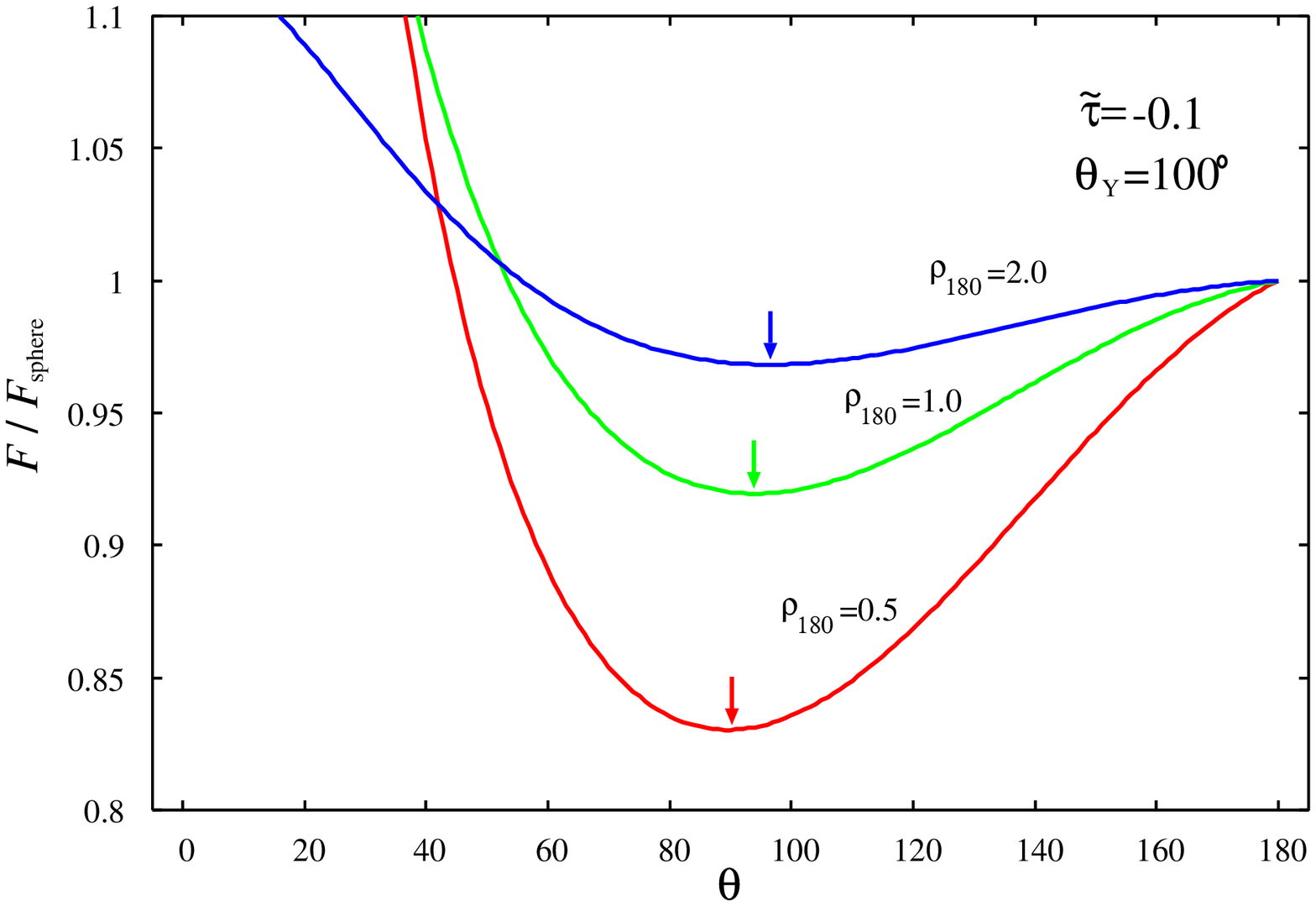}
\label{fig:S10b}
}
\end{center}
\caption{
The Helmholtz free energy landscape defined by Eq.~(\ref{eq:S4}) for a negative line tension $\tilde{\tau}=-0.1$ when (a) $\theta_{\rm Y}=40^{\circ}$ and (b) $\theta_{\rm Y}=100^{\circ}$.  The global minimum (down arrow) in (a) becomes deeper compared to Fig.~\ref{fig:S3}, and always represents a cap-shaped droplet.  Neither wetting nor drying occurs as the negative line tension always deepen the global minimum of the free-energy landscape except at $\theta=0^{\circ}$ and $\theta=180^{\circ}$.
 } 
\label{fig:S10}
\end{figure}

The effects of a negative line tension for a droplet on a spherical substrate are different from the effects of a positive line tension.  Figure~\ref{fig:S10} shows the free energy landscape for $\tilde{\tau}=-0.1$ when (a) $\theta_{\rm Y}=40^{\circ}$ and (b) $\theta_{\rm Y}=100^{\circ}$.  The equilibrium morphology that corresponds to the global minimum of the free energy is always a cap-shaped droplet.  Neither the wetting state of a wrapped droplet with $\theta_{\rm e}=0^{\circ}$ nor the drying state of a spherical droplet with $\theta_{\rm e}=180^{\circ}$ can appear.  This fact can be easily understood by comparing Fig.~\ref{fig:S3} and Fig.~\ref{fig:S10}(a). Since the last term of Eq.~(\ref{eq:S4}) is always negative and lowers the free-energy curve when $\tilde{\tau}<0$ except at $\theta=0^{\circ}$ and $180^{\circ}$, where the last term vanishes, the free-energy minimum becomes lower than that when $\tilde{\tau}=0$ and will never exceed the one at $\theta=0^{\circ}$ and $\theta=180^{\circ}$. 

\begin{figure}[htbp]
\begin{center}
\subfigure[]
{
\includegraphics[width=0.9\linewidth]{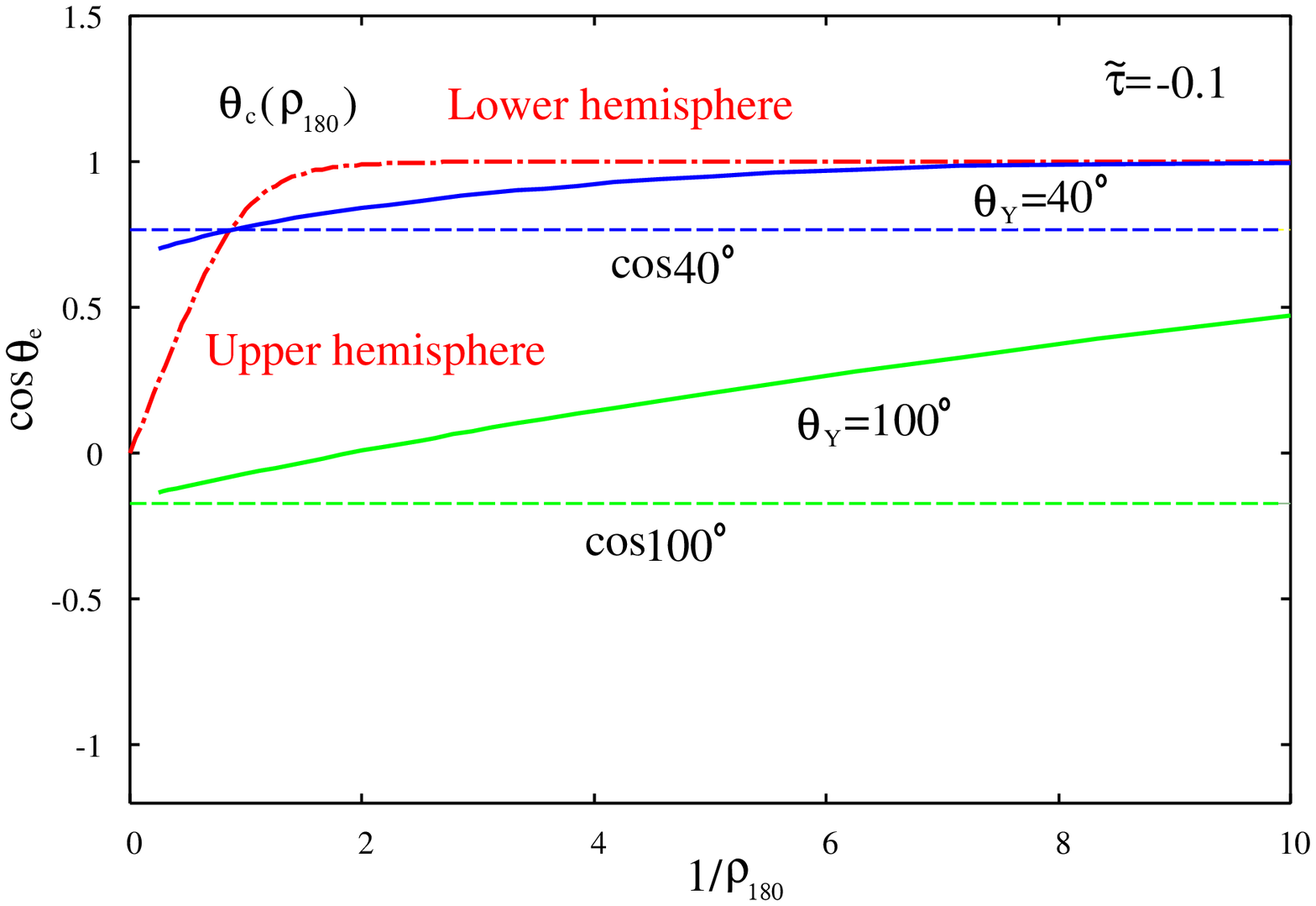}
\label{fig:S11a}
}
\subfigure[]
{
\includegraphics[width=0.9\linewidth]{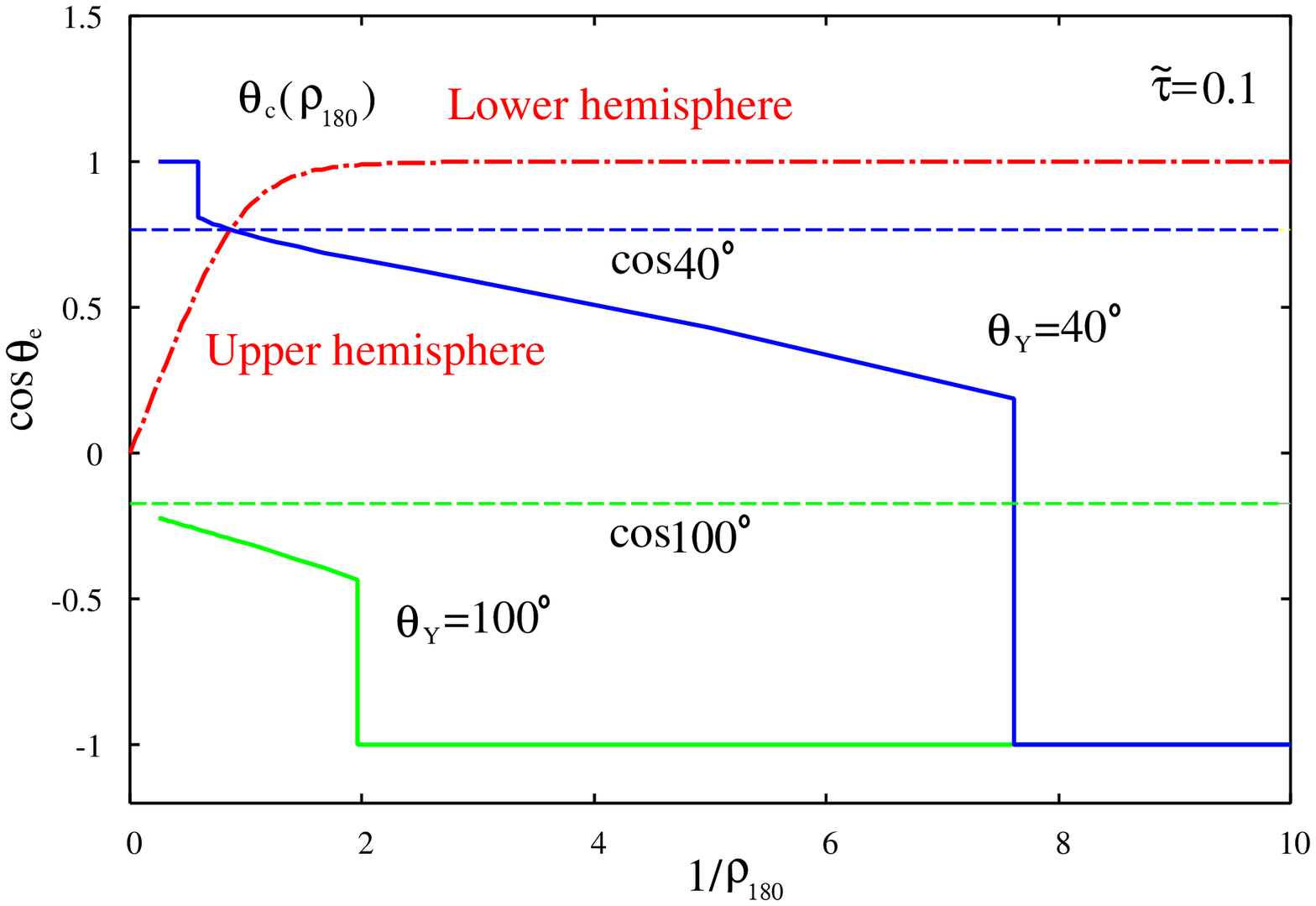}
\label{fig:S11b}
}
\end{center}
\caption{
The cosine of the equilibrium contact angle $\cos\theta_{\rm e}$ for (a) a negative ($\tilde{\tau}=-0.1$) and (b) a positive ($\tilde{\tau}=+0.1$) line tension (c.f. Fig.~\ref{fig:S6}) as a function of the inverse of the droplet size $1/\rho_{180}$ when the substrate is hydrophilic ($\theta_{\rm Y}=40^{\circ}$) and hydrophobic ($\theta_{\rm Y}=100^{\circ}$).  The dot-dash curve represents the cosine of the characteristic contact angle $\cos\theta_{\rm c}$ defined by Eq.~(\ref{eq:S12}) as a function of $1/\rho_{180}$.  The equilibrium contact angles for the negative line tension in (a) are increasing functions (two solid curves), while the ones for the positive line tension in (b) are decreasing functions (two solid curves). The metastable branches in Fig.~\ref{fig:S6} are omitted.
 } 
\label{fig:S11}
\end{figure}

Furthermore, $\cos\theta_{\rm e}$ {\it increases} as the $1/\rho_{180}$ increases when $\tilde{\tau}=-0.1$ as shown in Fig.~\ref{fig:S11}(a), while $\cos\theta_{\rm e}$ {\it decreases} as $1/\rho_{180}$ increases when $\tilde{\tau}=+0.1$ as shown in Fig.~\ref{fig:S11}(b).  Also, the nonlinearity of $\cos\theta_{\rm e}$ is more pronounced for the negative line tension. This behavior has already been predicted on flat substrates since Eq.~(\ref{eq:S11}) becomes
\begin{equation}
\cos\theta_{\rm e}=\cos\theta_{\rm Y}-\frac{\tilde{\tau}}{\rho_{\rm e}\sin\theta_{\rm e}}
\approx \cos\theta_{\rm Y}-\frac{\tilde{\tau}}{\rho_{180} }
\label{eq:S22}
\end{equation}
since $R\rightarrow\infty$ ( $\rho_{\rm e}\rightarrow 0$), where $\rho_{\rm e}\sin\theta_{\rm e}$ is the contact line radius of the droplet on a flat substrate,  and this behavior was actually observed experimentally~\cite{Pompe2000,Wang2001,Checco2003}.  Therefore, a similar behavior is predicted and is actually observed in Fig.~\ref{fig:S11} for a spherical substrate from Eq.~(\ref{eq:S11}).   Hence, a hydrophobic spherical surface becomes {\it less} hydrophobic and a hydrophilic surface becomes {\it more} hydrophilic for positive line tensions when the droplet volume is increased, while, a hydrophobic surface becomes {\it more} hydrophobic and a hydrophilic surface becomes {\it less} hydrophilic for negative line tensions. Therefore, the appearance of a drying spherical droplet or a wetting wrapped spherical droplet on a spherical substrate (Fig.~\ref{fig:S7}) may indicate the existence of a positive line tension, though the contact angle hysteresis cannot be ruled out.

The capillary model of the cap-shaped droplet employed in this work possesses short-wavelength instability~\cite{Dobbs1999,Brinkmann2005,Guzzardi2006} for negative line tension on a flat substrate because the undulation of the contact line around the circular shape increases the contact-line length and decreases the free energy.  Therefore, a cap-shaped droplet with a negative line tension may not be stable.  However, Mechkov et al.~\cite{Mechkov2007} showed that this is not a physical instability when the molecular interaction near the three-phase contact line is included using the disjoining pressure and the interface-displacement model~\cite{Indekeu1992,Napari2003,Boinovich2011}.  Furthermore, Berg et al.~\cite{Berg2010} argued that the instability due to the negative line tension is stabilized by a higher-order term of curvature. The result for the negative line tension in Fig.~\ref{fig:S10} is simply a mathematical consequence within the capillary model without fluctuation.

The contact angle hysteresis is also neglected.  We did not consider the fluctuation of line tension $\pm \Delta\tau$ caused by the local impurity or defect, which will leads to the contact angle hysteresis between advancing and receding contact angles~\cite{Tadmor2004}.  Instead, we considered that the line tension and the generalized Yong's equation give the observed equilibrium contact angle.  The contact angle hysteresis could be superimposed on the volume dependent contact angle in Figs.~\ref{fig:S6} and \ref{fig:S11}.  

It is possible to study the morphological phase transition when the wettability represented by the Young's contact angle $\theta_{\rm Y}$ is altered, for example, by electrowetting~\cite{Mugele2005} for a fixed droplet volume.  In this case, we can change the Young's contact angle along a vertical line instead of on the droplet radius $\rho_{180}$ along the horizontal line in Fig.~\ref{fig:S4} (see also Fig.~7 and Fig.~12 of the reference \cite{Iwamatsu2016b}).  In this case, we may always encounter the wetting transition as well as the drying transition by changing the Young's contact angle $\theta_{\rm Y}$, and we can consider a situation similar to the one when the droplet volume is altered on  a hydrophilic substrate. 

Finally, we describe the experimental verification of our theoretical prediction and the limitation of our theoretical model.  First, the line tension must be as large as $\tilde{\tau}\sim 0.1$.  Suppose the liquid has a high surface tension $\sigma_{\rm lv}\sim 70 {\rm mNm}^{-1}$ (water) and the line tension is as small as~\cite{Schimmele2007} $\tau\sim 10^{-11}{\rm N}$, we have $R=\tau/\left(\tilde{\tau}\sigma_{\rm lv}\right) \sim 10^{-10}{\rm m}$.  Therefore, we would need a nanometer sized spherical substrate to observe the wetting and the drying with an atomic force microscope (AFM).  Typical optical microscope measurements~\cite{Gaydos1987,Drelich1996,Tao2011} may not be able to detect the volume dependence of the contact angle.  To observe the line-tension effect in macro- and micro-scale droplets, much larger~\cite{Drelich1996} line tension of the order of $\tau\sim 10^{-5}-10^{-6}{\rm N}$ would be necessary.  Recently, however, an ultra-low surface tension $\sigma_{\rm lv}\sim 10^{-7} {\rm N/m}$ and a line tension $\tau\sim 10^{-12}-10^{-13}{\rm N}$ have been predicted for a colloid-polymer mixture~\cite{Vandecan2008}.  Then, the size of the substrate can be  as large as $R \sim 10^{-5}-10^{-6}{\rm m}$. The energy barrier to induce wetting and drying transitions are at most of the order of $F/F_{\rm sphere}\sim 0.1$ from Fig.~\ref{fig:S5}(a).  The necessary energy will be as small as $\Delta F\sim 10^{-17} {\rm J}$ from Eq.~(\ref{eq:S3}) for $R=10^{-5} {\rm m}$, and $\rho=1$, which can be easily supplied by external stimuli such as pressure or vibrations.  It could be possible to observe a morphological transition using an optical microscope~\cite{Drelich1996} on a micro-scale substrate and a droplet of colloid-polymer mixture instead of molecular liquids.

The above discussion and our model assumed that the line tension $\tau$ is constant and does not depend on the size and the contact angle of the droplet.  In fact, it is well recognized that the magnitude of the line tension depends on the size of the droplet.  A larger droplet has a larger line tension, which can be as large as $\tau\sim 10^{-5}{\rm N}$~\cite{David2007}.  The line tension can be constant only when the radius of three-phase contact line is less than $10^{-7}{\rm m}$~\cite{Heim2013}. In fact, the line tension should depend not only on the size of droplet but also on the contact angle~\cite{David2007,Heim2013}, since intermolecular forces at the three-phase contact line are affected by the geometry at the contact line.  Then, we cannot use the generalized Young's equation Eq.~(\ref{eq:S11}), which is derived from the variation under the condition of constant line tension~\cite{Schimmele2007}.  Furthermore, our model used Helmholtz free energy for a fixed volume.  The volume change of the droplet is controlled not by the surrounding vapor pressure but by the forced injection (extraction) of the liquid.  Therefore, our model is appropriate only for non-volatile liquids, and cannot be used to describe the size dependence over several orders of magnitude.

\section{\label{sec:sec4}Conclusion}

In this study, we considered the size-dependent contact angle and the wetting and drying transition of a spherical cap-shaped droplet placed on a spherical substrate when the line tension effect is included within the capillary model.  The contact angle is determined from the generalized Young's equation, which includes the effects of line-tension.  The morphology of the droplet is studied using the mathematically rigorous formula for the Helmholtz free energy.  The morphological drying transition from a cap-shaped droplet to a spherical droplet and the wetting transition from a cap-shaped droplet to a wrapped spherical droplet were predicted for a positive line tension.   The scenarios for these morphological transitions were deduced from the free-energy landscape of the Helmholtz free energy. The contact angle is a decreasing function with respect to the droplet size when the line tension is positive, while the angle is an increasing function of the droplet size when the line tension is negative.  These morphological transitions and the size-dependent contact angle by line tension could be observed for nano-meter sized droplets and substrates in case of the small line tension observed in those nano-droplets.  However,  if it is possible to realize the large line tensions and the very low liquid-vapor surface tension, morphological transitions and size-dependent contact angle in mocro-meter sized droplets may be observed.



\end{document}